\documentclass[%
 reprint,
superscriptaddress,
 amsmath,amssymb,
 aps,
prl,
]{revtex4-1}

\usepackage{graphicx}% Include figure files
\usepackage{dcolumn}% Align table columns on decimal point
\usepackage{bm}% bold math
\usepackage{chemformula}
\usepackage{hyperref}% add hypertext capabilitie

\usepackage{color}

\newcommand{\angstrom}{\text{\normalfont\AA}}

\begin{document}

\preprint{APS/123-QED}

\title{Squeezing N\'eel-type Magnetic Modulations by Enhanced Dzyaloshinskii-Moriya interaction of $4d$ Electrons}

\author{{\'A}d{\'a}m Butykai}
 \email{butykai@mail.bme.hu}
\affiliation{Department of Physics, Budapest University of Technology and 
Economics 1111 Budapest, Hungary}
\affiliation{Condensed Matter Research Group of the Hungarian Academy of 
Sciences, 1111 Budapest, Hungary}

\author{Korbinian Geirhos} 
\affiliation{Experimental Physics V, Center for Electronic Correlations and Magnetism, University of Augsburg, 86135 Augsburg, Germany}

\author{D\'avid Szaller}%
\affiliation{Department of Physics, Budapest University of Technology and 
Economics 1111 Budapest, Hungary}
\affiliation{Institute of Solid State Physics, TU Wien, 1040 Vienna, Austria}

\author{L{\'a}szl{\'o} F. Kiss}
\affiliation{Department of Experimental Solid State Physics, Institute for Solid State Physics and Optics, Wigner Research Centre for Physics, 1121 Budapest, Hungary}

\author{L\'aszl\'o Balogh}%
\affiliation{Department of Physics, Budapest University of Technology and Economics 1111 Budapest, Hungary}

\author{Maria Azhar}%
\affiliation{Institut f\"ur Theoretische Festk\"orperphysik, Karlsruhe Institute of Technology, 76131 Karlsruhe, Germany}

\author{Markus Garst}%
\affiliation{Institut f\"ur Theoretische Festk\"orperphysik, Karlsruhe Institute of Technology, 76131 Karlsruhe, Germany}
\affiliation{Institute for Quantum Materials and Technology, Karlsruhe Institute of Technology, 76131 Karlsruhe, Germany}

\author{Lisa DeBeer-Schmitt}%
\affiliation{Neutron Scattering Division, ORNL, Oak Ridge, TN 37831, USA}

\author{Takeshi Waki}
\affiliation{Department of Materials Science and Engineering, Kyoto University, Kyoto 606-8501, Japan}

\author{Yoshikazu Tabata}
\affiliation{Department of Materials Science and Engineering, Kyoto University, Kyoto 606-8501, Japan}

\author{Hiroyuki Nakamura}
\affiliation{Department of Materials Science and Engineering, Kyoto University, Kyoto 606-8501, Japan}

\author{Istv{\'a}n K{\'e}zsm{\'a}rki}
\affiliation{Department of Physics, Budapest University of Technology and Economics 1111 Budapest, Hungary}
\affiliation{Experimental Physics V, Center for Electronic Correlations and Magnetism, University of Augsburg, 86135 Augsburg, Germany}

\author{S{\'a}ndor Bord{\'a}cs}%
 \email{bordacs.sandor@wigner.bme.hu}
\affiliation{Department of Physics, Budapest University of Technology and 
	Economics 1111 Budapest, Hungary}
\affiliation{Hungarian Academy of Sciences, Premium Postdoctor Program, 1051 
	Budapest, Hungary}

\date{\today}% It is always \today, today,

\begin{abstract}
In polar magnets, such as GaV$_4$S$_8$, GaV$_4$Se$_8$ and VOSe$_2$O$_5$, modulated magnetic phases namely the cycloidal and the Néel-type skyrmion lattice states were identified over extended temperature ranges, even down to zero Kelvin. Our combined small-angle neutron scattering and magnetization study shows the robustness of the N\'eel-type magnetic modulations also against magnetic fields up to 2\,T in the polar GaMo$_4$S$_8$. In addition to the large upper critical field, enhanced spin-orbit coupling produces a variety of modulated phases with sub-10\,nm periodicity and a peculiar distribution of the magnetic modulation vectors. Thus, our work demonstrates that non-centrosymmetric magnets with $4d$ and $5d$ electron systems are ideal candidates to host highly compressed magnetic spirals and skyrmions.  
\end{abstract}

\pacs{Valid PACS appear here}% PACS, the Physics and Astronomy
\maketitle

%\tableofcontents

%================================================================================================

\section{Introduction}

In the presence of strong spin-orbit coupling (SOC) topologically non-trivial states of condensed matter emerge such as the surface states of topological insulators \cite{Hasan2010TIreview,Qi2011TIreview}, Dirac and Weyl fermions \cite{Liu2014Discovery,Xu2015Discovery} or Majorana particles \cite{Rokhinson2012Fractional,Mourik2012Signature}. In spin systems the first order manifestation of the SOC is the antisymmetric Dzyaloshinskii-Moriya interaction (DMI), which gives rise to spin spirals and skyrmion lattice (SkL) states in non-centrosymmetric compounds \cite{dzyaloshinskii1964theory,bak1980theory,bogdanov1989thermodynamically}. The non-trivial topology of skyrmions \cite{nagaosa2012gauge,neubauer2009topological}, as well as their interaction with conduction electrons \cite{jonietz2010spin}, electric fields \cite{adams2012long,seki2012observation,seki2012magnetoelectric,White2014ElectricField} and spin-waves \cite{onose2012observation,seki2016magnetochiral,garst2017collective} motivated intense research exploring possible applications in next-generation data storage and microwave-frequency spintronic devices \cite{zhang2015skyrmion,parkin2015memory,hanneken2015electrical,maccariello2018electrical}.

The SkL phase was first observed in the chiral cubic helimagnet MnSi with moderate SOC \cite{muhlbauer2009skyrmion}. In cubic helimagnets the symmetry-dictated form of DMI enables Bloch-type magnetic modulations, i.e.~spin helices and Bloch-skyrmions. This Bloch-type SkL, comprising $q$-vectors perpendicular to the direction of the applied field, is stabilized by thermal fluctuations over the energetically favoured longitudinal conical state, only in the close vicinity of the Curie temperature \cite{muhlbauer2009skyrmion, bauer2016generic}. Cubic magnetocrystalline anisotropies that are higher-order terms in the SOC determine the orientation of helical order at zero field and induce small deflections of the SkL planes for fields applied along low-symmetry crystallographic directions \cite{bauer2016generic, adams2018response}. The increasing strength of the SOC was studied in Mn(Si$_{1-x}$Ge$_x$) by replacing Si with heavier Ge \cite{Fujishiro2019topological}. As a result, the periodicity of the magnetic modulation decreases and the SkL state is transformed to a hedgehog-lattice state. The enhanced cubic anisotropy can also lead to SkL states at low-temperatures, without relying on stabilization by thermal fluctuations \cite{Qian2018New,Chacon2018Observation}.

\begin{figure}
	\centering
	\includegraphics[width=\columnwidth]{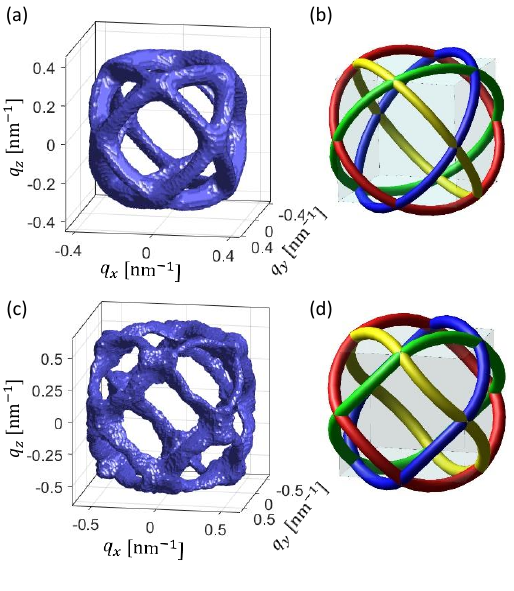}
	\caption{\textbf{A comparison of the reciprocal-space structure of the modulation wavevectors in GaV$_4$S$_8$ and GaMo$_4$S$_8$ at 12\,K and 2\,K, respectively.} The first row contains the SANS tomographic image and its graphical representation in GaV$_4$S$_8$ in panels (a) and (b), respectively. The scattering pattern contains four rings of $q$-vectors represented by distinct colors in panel (b), each corresponding to one of the rhombohedral domain states. Panels (c)-(d): Distribution of the magnetic wavevectors observed in GaMo$_4$S$_8$. The rings are deflected from the $\left\{111\right\}$-type planes in the segments between the $\left<110\right>$ directions in an alternating manner.}
	\label{fig:SANS_GVS_vs_GMS}
\end{figure}

For potential memory and spintronic applications a further reduction in the skyrmion size is desired, which can be achieved through the enhancement of the SOC. In the vast majority of the skyrmion host crystals reported to date, the magnetism is governed by the $3d$ electrons of transition metals, such as V, Mn, Fe, Co, Cu. Here, we demonstrate the emergence of particularly robust modulated magnetic phases  in the $4d$ polar magnet, GaMo$_4$S$_8$ \cite{rastogi1987magnetic, Powell2007Cation, neuber2018architecture, geirhos2018orbital, Geirhos2021Review}. In the rhombohedral phase, our small-angle neutron scattering (SANS) data (Fig.~\ref{fig:SANS_GVS_vs_GMS} (c)) show that the strong SOC reduces the periodicity of the magnetic structure to $\lambda\approx9.8$\,nm, being one of the shortest modulation observed in bulk crystals hosting the SkL state due to DMI. Moreover, it modifies the distribution of the $q$-vectors, as illustrated in Fig.~\ref{fig:SANS_GVS_vs_GMS} (d), which we explain by an effective Landau theory containing a cubic anisotropy term in addition to the axial anisotropy, inherent to the rhombohedral state. Our temperature, magnetic field and angular dependent magnetization and electric polarization measurements reveal a complex phase diagram where we attributed phases to the cycloidal and SkL states and observed a third modulated magnetic phase. Moreover, we found that the modulated spin states extend up to fields as high as 2\,T, which further support their robustness.

Spin cycloids and Néel SkL have been found recently in the polar phase of the lacunar spinels GaV$_4$S$_8$ and GaV$_4$Se$_8$  \cite{kezsmarki2015neel,white2018direct,bordacs2017equilibrium,fujima2017thermodynamically}, which are narrow-gap multiferroic semiconductors \cite{pocha2000electronic, Reschke2020Phonons}. The polar rhombohedral structure (space group $R3m$) develops via a cooperative Jahn-Teller distortion driven by the unpaired electron of each V$_4$ cluster with spin S=1/2 occupying a triply degenerate orbital in the high-temperature cubic state ($F\overline{4}3m$) \cite{pocha2000electronic}. The rhombohedral distortion can occur as elongation of the unit cell along any of the four $\left<111\right>$-type cubic directions, thus, four rhombohedral domain states, that we denote by P$_{1-4}$, are present below the structural transition at $\sim$42\,K \cite{pocha2000electronic,ruff2015multiferroicity,bordacs2017equilibrium}. In contrast to the chiral cubic helimagnets, in these lacunar spinels the cycloidal character of the modulations with $q$-vectors restricted to the plane perpendicular to the rhombohedral axis (see Fig.~\ref{fig:SANS_GVS_vs_GMS} (a)-(b)) enhances the stability range of the SkL phase \cite{kezsmarki2015neel,bordacs2017equilibrium,leonov2017skyrmion}. In both compounds, small-angle neutron scattering (SANS) experiments revealed magnetic modulations with wavelengths of $\lambda\sim$20\,nm \cite{kezsmarki2015neel,white2018direct,bordacs2017equilibrium}, indicating a similar ratio of the exchange interactions and the DMIs. 

In the compound GaMo$_4$S$_8$ studied here, the tetrahedral Mo$_4$ clusters carry an unpaired hole with spin S=1/2 \cite{pocha2000electronic}. Correspondingly, the cubic state is rhombohedrally distorted by the compression of the unit cell along one of the $\left<111\right>$-type directions \cite{Powell2007Cation}. Although, Rastogi et al., pointed out the importance of the correlation in GaMo$_4$S$_8$ and found a rich magnetic phase diagram below $T_C=19$\,K long ago \cite{rastogi1987magnetic}, the spin ordering patterns of these phases have not been studied.

%===============================================================================

\section{Results}

\begin{figure*}
	\centering
	\includegraphics[width=\textwidth]{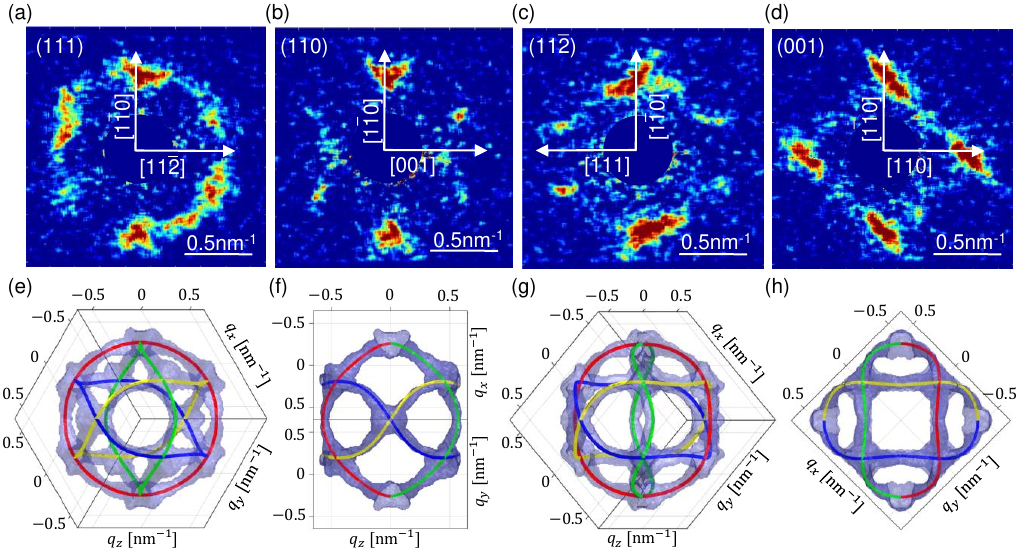}
	\caption{\textbf{SANS images of GaMo$_4$S$_8$ recorded at $T=$2\,K (panels (a)-(d)) and reciprocal space distribution of the cycloidal wavevectors (panels (e)-(h)).} In panels (e)-(h) the q-vector distribution is	shown from the direction of the neutron beam in (a)-(d), respectively. The perturbative solution of the model potential in Eq.~(\ref{eq:Qmodel}) fitted to the experimental data is visualized by solid curves plotted over the experimental data. The four different colors represent scattering from the four rhombohedral domain states.}
	\label{fig:GVS_RST_highsymm}
\end{figure*}

\subsection{Wavevector distribution of the magnetic cycloid at zero field}

We explored the zero-field modulated magnetic states of GaMo$_4$S$_8$ by SANS experiments. The scattered intensity was recorded in zero-field at $T=2$\,K upon the 180$^{\circ}$ rotation of the sample in 1$^{\circ}$ steps, with an acquisition time of 120\,s at each angle. The background signal was measured in the paramagnetic phase at $T=25$\,K following the same procedure. The scattering images were averaged over a 10$^{\circ}$ moving window in the rotation angle to improve the signal-to-noise ratio. Figures \ref{fig:GVS_RST_highsymm} (a)-(d) show the SANS images obtained on four high-symmetry planes, namely the (111), (110), $\left(11\bar{2}\right)$ and (001) planes. A pixel-wise adaptive Wiener filter, assuming Gaussian noise, was applied for better visualization. 

The q-dependence of the scattering intensity in the (111) plane, averaged over the polar angles, was fitted by a Gaussian, yielding $\left|\mathbf{q}\right| = 0.64$\,nm$^{-1}$ for the length of the modulation vectors with a FWHM of 0.2\,nm$^{-1}$, which corresponds to a real-space periodicity of $\lambda\approx 9.8$\,nm. The uncertainty of $|q|$ mainly originates from the broad and anisotropic distribution of the scattering intensity. Whereas the Curie-temperature is close to that of GaV$_4$Se$_8$, implying a similar strength of the symmetric exchange ($J$) in the two compounds, the modulation wavelength in GaMo$_4$S$_8$ is roughly half of that in GaV$_4$S$_8$ \cite{kezsmarki2015neel} and GaV$_4$Se$_8$ \cite{bordacs2017equilibrium}, implying a stronger DMI coupling ($D$), as $\lambda \propto J/D$. 

Over the wide-angle rotation experiment, each scattering image represents a planar cross section of the three-dimensional distribution of the $q$-vectors, where the azimuthal angle of the vertical slicing plane is varied via stepwise rotation of the sample. The 3D scattering pattern was reconstructed using the whole set of the cross section images. In order to enhance the signal-to-noise ratio and to eliminate the asymmetries of the scattering pattern introduced by imbalances between the populations of the different structural domain states, the 3D scattering pattern was symmetrized for all the symmetry operations of the cubic $T_d$ point group \footnote{for details, see Supplementary Information}. Figures~\ref{fig:GVS_RST_highsymm} (e)-(h) display the symmetrized image as viewed from the different high-symmetry directions. 

It is instructive to compare the reciprocal-space $q$-distributions in GaV$_4$S$_8$ and GaMo$_4$S$_8$, as shown in Figs.~\ref{fig:SANS_GVS_vs_GMS} (a) and (c), respectively. The neutron scattering data collected in the zero-field cycloidal phase of GaV$_4$S$_8$ at 12\,K is reproduced from Ref.~\onlinecite{white2018direct}.  In both compounds the cycloidal $q$-vectors are distributed over four intersecting rings corresponding to the four structural domain states. The ring structure, instead of six well-defined Bragg spots, is due to static orientational disorder of the q-vectors, as discussed in Ref.~\onlinecite{white2018direct}. However, in contrast to GaV$_4$S$_8$, where the modulation vectors are evenly distributed over rings restricted to the $\left\{111\right\}$-type planes, in GaMo$_4$S$_8$, the four rings of the $q$-vectors wave out of the $\left\{111\right\}$ planes, crossing them only along the $\left<110\right>$-type directions. This waving pattern of the $q$-vectors preserves the three-fold rotational symmetry of the rhombohedral structure as highlighted by the schematic image in Fig. \ref{fig:SANS_GVS_vs_GMS} (d). 

%================================================================================================
%Fitting
To explain the distribution of the $q$-vectors, the following effective 
Landau potential for the unit vector $\hat{q}$ is considered with its $x$,$y$,$z$ components defined in the 
cubic setting,
\begin{equation}
\mathcal{V}(\hat{q})=(\hat{n}\hat{q})^2+\alpha\left(\hat{q}_x^4+\hat{q}_y^4+\hat{q}_z^4\right)+....
\label{eq:Qmodel}
\end{equation}
The first term describes the uniaxial anisotropy emerging in the rhombohedral phase with the $\hat{n}$ unit vector parallel to any of the four $\left<111\right>$-type polar axes, and the second term is the lowest-order term compatible with the cubic T$_d$ symmetry. The length of the $q$-vectors is essentially fixed by the DMI, $q \sim D/J$, which is consistent with the experimental SANS data. The coefficient $\alpha$ parameterizing the relative strengths of the two terms in Eq.~(\ref{eq:Qmodel}) is thus effectively of second order in the SOC. The first term favours the confinement of the $q$-vectors normal to the polar axes, $\hat{n}$, as imposed by the DMI. The waving of the $q$-vectors out of the \{111\} planes is captured by the second term. On the microscopic level it represents magnetocrystalline contributions to the Ginzburg-Landau theory for the magnetization that are effectively of fourth order in the SOC. 

The minimal-energy solutions to Eq.~\eqref{eq:Qmodel} are sought by parametrizing $\hat{q}$ in spherical coordinates. The polar angle, $\Theta$ is chosen with respect to the polar axis $\hat{n}\parallel\left<111\right>$ of each domain. The azimuthal angle, $\Phi$ is enclosed between the in-plane component of $\hat{q}$ and one of the corresponding  
$\left<1\bar{1}0\right>$ directions.
%$\Theta$ and $\Phi$, on the surface of the unit sphere, where the orthonormal coordinate system $\{\hat{\mathbf{e}}_1, \hat{\mathbf{e}}_2,\hat{\mathbf{e}}_3\}$ is chosen for each domain state such that $\hat{\mathbf{e}}_3$ is parallel with the polar axis $\hat n$, whereas $\hat{\mathbf{e}}_1$ and $\hat{\mathbf{e}}_2$ are parallel to the corresponding $\left<1\bar{1}0\right>$ and $\left<11\bar{2}\right>$-type directions normal to the polar axis. 
%
In the limit of weak SOC, $\alpha$ is negligible and the wavevectors are basically in-plane ($\Theta \approx \pi/2$), since the spirals are degenerate with respect to the azimuthal angle $\Phi$. This gives rise to an equal distribution on circles, as observed for GaV$_4$S$_8$ in Fig.~\ref{fig:SANS_GVS_vs_GMS}(a). Expanding the potential in $\Theta$ around $\pi/2$ up to the second order and minimizing one obtains for the deviations from the circle
\begin{equation}
\Theta(\Phi)=\frac{\pi}{2}+\frac{\sqrt{2}}{3}\frac{\alpha}{1+\alpha}\sin{(3\Phi)},
\label{eq:qpert}
\end{equation}
that describes the waving of $\hat q$ in GaMo$_4$S$_8$.  The least-square fitting of the data to Eq.~\eqref{eq:qpert} yields $\alpha=-0.14 \pm 0.003$ and $|\mathbf{q}|=0.64$\,nm$^{-1} \pm 10^{-3}$ ($\lambda=9.81$\,nm). As shown in 
Figs.~\ref{fig:GVS_RST_highsymm} (e)-(h), the fitted model is in excellent 
agreement with the SANS data, indicating the significant influence of cubic 
anisotropies in GaMo$_4$S$_8$ as opposed to GaV$_4$S$_8$,\footnote{For GaV$_4$S$_8$ we find within the experimental precision $\alpha=0 \pm 0.04$. The uncertainty is mainly determined by the image preconditioning rather than the accuracy of the fitting, for details see SI.} in accord with the stronger atomic SOC of Mo. 

The tilting of $\hat q$ out of the \{111\} planes as well as higher order SOC represented by additional terms $\hat q_x^6 + \hat q_y^6 + \hat q_z^6$ in the Landau potential break the degeneracy of q-vectors around the ring and favour either $\langle 1\bar 1 0\rangle$ or $\langle 11\bar 2\rangle$-type directions. However, within our experimental accuracy we were not able to resolve an enhanced intensity along any of these directions, see Supplementary information.

%================================================================================================
%Magnetic Phase Diagram

\subsection{Magnetic phase diagram at finite magnetic fields}

\begin{figure}
	\centering
	\includegraphics[width=\columnwidth]{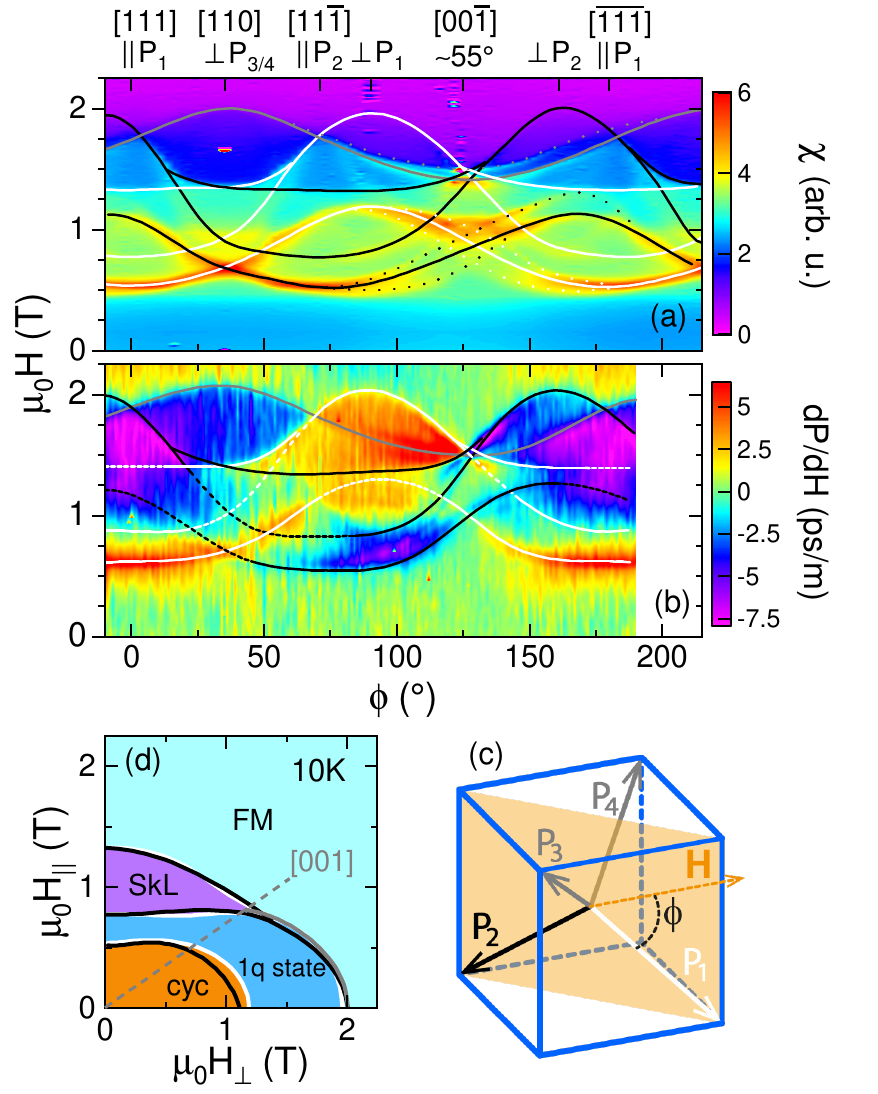}
	\caption{\textbf{Angular dependence of the magnetic phase diagram at 10\,K.}	Angular dependence of the magnetic susceptibility and magnetocurrent measured at 10\,K are displayed in panel (a) and (b), respectively. Panel (c) illustrates the orientations of the four polar axes (P$_{1-4}$) and the plane where the magnetic field was rotated by angle $\phi$. The lines represent the magnetic phase boundaries for the different domains (white: P$_1$, black: P$_2$, grey: P$_3$ and P$_4$). Dotted lines in panel (a) represent a splitting of the phase boundary, while dashed lines in panel (b) indicate regions where the anomalies could not unambiguously be identified. The magnetic phase diagram constructed based on the angular dependence of these anomalies at 10\,K is shown in panel (d). cyc, SkL and 1q state respectively label the cycloidal, skyrmion lattice and a modulated phase described by a single q-vector. H$_\parallel$ and H$_\perp$ are field components parallel and perpendicular to the polar axis, respectively.} 
	\label{fig:GMS_PD_Rot}
\end{figure}

We explore the modulated magnetic states stabilized by the interplay of external magnetic fields and the strong SOC of GaMo$_4$S$_8$. Due to the polar symmetry of this compound the phase diagram may depend on the angle between the polar axis and the field, thus, we measured longitudinal magnetic and differential magnetoelectric susceptibility at 10\,K while the field was rotated in finite steps within the (1$\bar 1$0) plane between successive field sweeps. The obtained data are respectively shown in Fig.~\ref{fig:GMS_PD_Rot} (a) and (b) as color maps over the field magnitude-orientation plane, where the angle $\phi$ was measured from the [111] axis as sketched in Fig.~\ref{fig:GMS_PD_Rot} (c). The critical fields identified as peaks or sharp steps in the susceptibility and magnetocurrent curves are indicated by lines in Fig.~\ref{fig:GMS_PD_Rot} (a) and (b). 

Compared to the early magnetization study performed on powder samples \cite{rastogi1987magnetic}, in the single crystal sample we resolved more than two phase transitions strongly depending on the direction of the field. The interpretation of the complicated angular dependent pattern of the anomalies can be simplified by assuming that all four polar domain states are present in the sample. Anomalies and their replica appearing at positions shifted by $\sim$109$^\circ$ are attributed to the P$_1$ (white lines) and the P$_2$ (black lines) domains since both polar axes lie within the rotation plane and they span $\sim$109$^\circ$. The remaining anomaly (grey line) should correspond to both the P$_3$ and P$_4$ domains, when the field is rotated in the high-symmetry (1$\bar 1$0) plane. The field direction [001] is even more special as the polar axes span the same 55$^\circ$ with the field in all domains, thus, the lines should intersect each other. Such an intersection occurs at $\sim$1.5\,T where the small deviations likely caused by a slight misorientation of the sample. However, in the same angular range phase boundaries are split around 0.75\,T indicated by dotted lines in Fig.~\ref{fig:GMS_PD_Rot} (a). We found that anomalies observed on the different domains collapse into a common phase diagram when plotted on the H$_\parallel$ - H$_\perp$ plane, where H$_\parallel$ and H$_\perp$ represents the field component parallel and perpendicular to the polar axis, respectively (see Fig.~\ref{fig:GMS_PD_Rot} (d)). This confirms that the sample is in a polar multidomain state and the magnetic state stabilized by the field depends only on the angle between the polar axis and the field. 
%this is questionable as the lower field phase boundary is not observed for P3 and P4
%The in-plane orientation of the field component respectively pointing along the $\langle 11\bar 2\rangle$ and $\langle 11 0\rangle$-type directions in P$_{1-2}$ and P$_{3-4}$ domains does not influence the magnetic phases. 
We note that the finite magnetocurrent signal detected in all phases implies that the modulated magnetic states couple to the electric polarization as observed in the sister compounds GaV$_4$S$_8$ and  GaV$_4$Se$_8$ \citep{ruff2015multiferroicity,fujima2017thermodynamically,Geirhos2020Macroscopic}. This magnetoelectric coupling is allowed by the polar symmetry of GaMo$_4$S$_8$ \citep{ruff2015multiferroicity}, and the magnetic field induced changes of the spin texture modify the polarization texture.

%Or maybe this is the reason why the grey line is missing for P$_{3-4}$?

 \begin{figure}
	\centering
	\includegraphics[width=\columnwidth,trim=0 0 120 0]{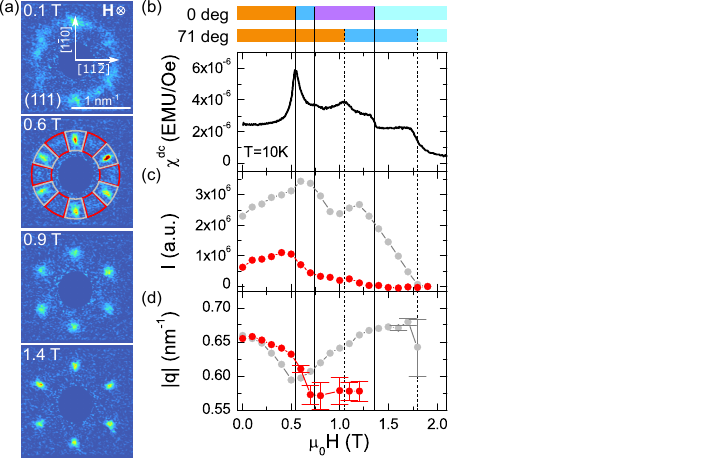}
	\caption{\textbf{Magnetic field dependence of the spin texture at 10\,K.} SANS images acquired at 10\,K in increasing magnetic fields are shown in panel (a). Both the neutron beam and the direction of the magnetic field was parallel to the [111] direction. The magnetic field dependence of the susceptibility shown in panel (b) is compared with the scattering intensity (panel (c)) and the position of the intensity peak (panel (d)) measured in the grey and red sections of the SANS patterns. The coloured stripes (same colour coding as in Fig.~\ref{fig:GMS_PD_Rot}) at the top of panel (b) represent the sequence of phase transitions in P$_1$ and P$_{2-4}$ domains with polar axes spanning 0$^\circ$ and $\sim$71$^\circ$ with respect to the field.}  
	\label{fig:GMS_SANS_Bdep}
\end{figure}

\begin{figure}
	\centering
	\includegraphics[width=\columnwidth]{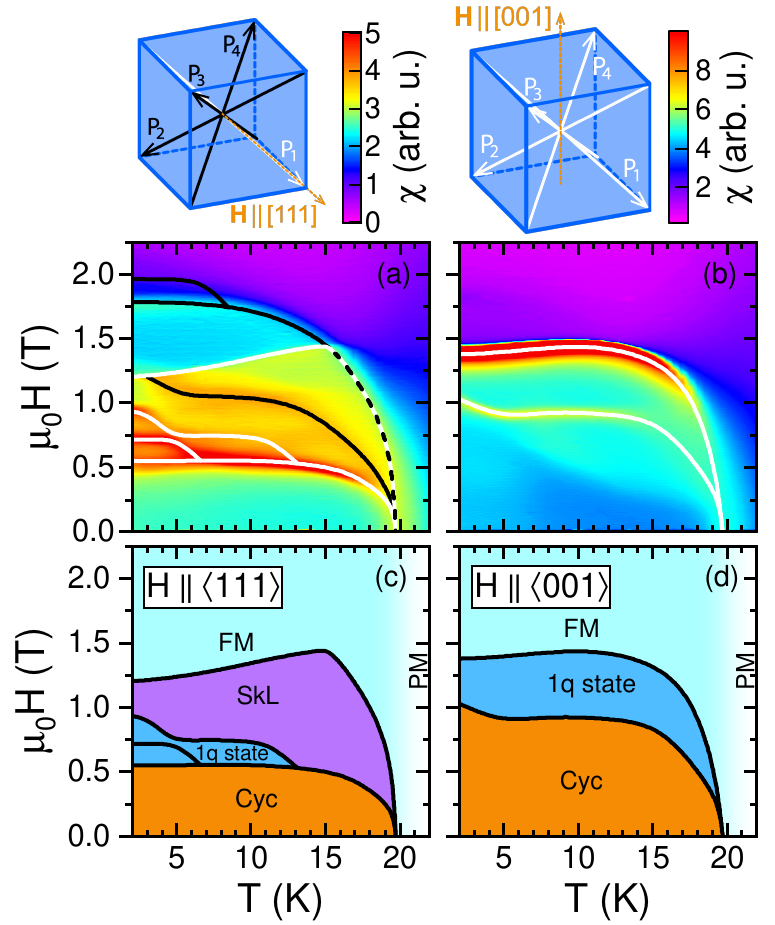}
	\caption{\textbf{Magnetic field and temperature dependence of the phase diagram for field parallel to the [111] and [001] directions.} The magnetic suscpetibility as a function of magnetic field and temperature is shown for field parallel to the [111] and [001] directions in panel (a) and (b), respectively. The direction of the magnetic field with respect to the four polar axes (P$_{1-4}$) is indicated above the color plots. The lines indicate anomalies of the susceptibility corresponding to phase transitions. For H$\parallel$[111] transitions occurring in the P$_1$ domain are highlighted by white lines whereas phase boundaries of the P$_{2-4}$ domains are drawn in black. For H$\parallel$[001] all four domains are indistinguishable. Panel (c) and (d) display schematic phase diagrams deduced for a polar monodomain sample for fields respectively spanning zero and 55$^\circ$ with the polar axis.}  
	\label{fig:GMS_PD_Tdep}
\end{figure}

In order to elucidate the nature of the various phases, we explored their properties for certain directions of the applied field in more details. We studied the magnetic field induced changes in spin textures at 10\,K by SANS experiments performed for fields parallel to the [111] axis on the (111) scattering plane. A representative set of SANS images are displayed in Fig.~\ref{fig:GMS_SANS_Bdep} (a). The smeared wavy ring of intensity present in low magnetic fields, see Fig.~\ref{fig:GVS_RST_highsymm} a, continuously evolves to well-defined Bragg spots in moderate fields pointing to a field-driven enhancement of the magnetic correlation length. To make our analysis more quantitative the field dependence of the scattering intensity and the average length of the $q$-vectors are plotted in Fig.~\ref{fig:GMS_SANS_Bdep} (c) and (d), respectively, in comparison with the susceptibility shown in Fig.~\ref{fig:GMS_SANS_Bdep} (b). The grey and red curves are obtained by fitting a Gaussian peak on the azimuthally averaged scattering intensity around the $\langle 1\bar1 0\rangle$ and $\langle 11\bar 2\rangle$-type directions highlighted by grey and red sectors in Fig.~\ref{fig:GMS_SANS_Bdep} (a). 

The phase transitions are marked by clear anomalies of the scattering parameters. As we have demonstrated above, only the P$_1$ domain, responsible for the red ring in Figs.~\ref{fig:SANS_GVS_vs_GMS} (d) \& \ref{fig:GVS_RST_highsymm}, scatters neutrons into the red sections whereas all domains contribute to the intensity detected in the grey regions. Remarkably, our measurements evidence that modulated magnetic structures are extremely robust against a magnetic field, i.e.~they extend up to 1.3\,T and 1.8\,T for fields parallel to and inclined at 71$^\circ$ from the polar axis, respectively. Such exceptional stability of modulated bulk phases has been observed so far only in centrosymmetric materials, hosting nearly atomic-scale skyrmions due to exchange frustration \cite{Kurumaji2019Skyrmion,Hirschberger2019Skyrmion}. As the modulated phases are expected to be stable up to a critical field of order $H_{FM} \propto D^2/J$, this finding corroborates that the DMI is the strongest in GaMo$_4$S$_8$ among the lacunar spinels known to host SkLs, likely due to enhancement of the SOC from $3d$ to $4d$ electrons. Interestingly, after a low-field decrease $|$q$|$ increases in higher fields in the grey region, which is unusual among skyrmion host spiral magnets. Since $|$q$|$ does not change above 1.3\,T, where only the P$_{2-4}$ domains contribute to the SANS intensity, the anomalous field induced shortening of the magnetic periodicity occurs in the P$_1$ domain, i.e.~for field parallel to the polar axis.

Since the DMI pattern compatible with the C$_{3v}$ symmetry allows cycloidal modulation we assign the zero field ground state of GaMo$_4$S$_8$ to a single-q cycloidal state (orange region in Fig.~\ref{fig:GMS_PD_Rot} (d)) in analogy with GaV$_4$S$_8$ and GaV$_4$Se$_8$. The hexagonal SANS pattern with minimal intensity in the red sections, that is observed above $\sim$0.75\,T, imply the formation of a SkL state or can alternatively manifest the coexistence of cycloidal domains with $q$-vectors pointing to the different $\langle1\bar{1}0\rangle$-type directions. Following the correspondence with the phase diagram of the sister compounds we rather attributed the purple phase pocket in Fig.~\ref{fig:GMS_PD_Rot} (d) to the SkL state. The robustness of this phase against strong tilting of the magnetic field from the polar axes implies an easy-axis type magnetic anisotropy, thus, it has the same character as in GaV$_4$S$_8$ \cite{leonov2017skyrmion,ehlers2016exchange}. Since the phase between the cycloidal and the field-polarized FM states for fields perpendicular to the polar axis (coloured in blue in Fig.~\ref{fig:GMS_PD_Rot} (d)) has a modulated spin structure possessing finite magnetization, it is likely to be a fan or a conical phase, which is not present in any other lacunar spinels. The stability range of this intermediate state narrows as the field is rotated toward the polar axis, however, it can separate the cycloidal and SkL phases even in parallel fields. The anomalous increase in $|$q$|$ above 0.5\,T also corresponds to the emergence of this single-q state.

The temperature-field phase diagram is mapped by susceptibility measurements below $T_C=19$\,K as shown in Fig.~\ref{fig:GMS_PD_Tdep} (a) and (b) for fields along [111] and [001] directions, respectively. Schematic phase diagrams deduced for a polar monodomain sample are presented in Fig.~\ref{fig:GMS_PD_Tdep} (c) and (d) for fields respectively spanning zero and 55$^\circ$ with the polar axis. The latter case, H$\parallel$[001] likely provides a representative phase diagram for the whole angular range 55$^\circ$-90$^\circ$ as evidenced for 10\,K in Fig.~\ref{fig:GMS_PD_Rot} (d). All phases observed at 10\,K including the SkL state extend down to the lowest temperatures, which suggests that the easy-axis anisotropy is smaller in GaMo$_4$S$_8$ compared to GaV$_4$S$_8$ \cite{white2018direct}. The conical phase enters between the cycloidal and the SkL phases only below 13\,K for fields applied along the polar axis, and its field stability range grows toward low temperatures. Moreover, below 6\,K additional anomalies of the susceptibility appear suggesting the emergence of a new phase not present at 10\,K in Fig.~\ref{fig:GMS_PD_Rot}.

\section{Discussion}

The main features of the phase diagram and their assignment is consistent with a recent theoretical study of GaMo$_4$S$_8$ that combines DFT calculations and Monte-Carlo simulations \cite{zhang2019possible}. This work predicts only two modulated structures, the cycloidal and the N\'eel-type SkL states for field parallel to the polar axis. With a moderate strength of anisotropy, these states were found to be stable down to the lowest temperatures, with a periodicity of $\sim$14.6\,nm , which is close to the experimentally observed one. However, further theoretical efforts are required to understand the emergence of the single-q state between the cycloidal and the SkL phases as well as the additional low-temperature phases. An important direction of future research is, in particular, the theoretical investigation of a micromagnetic model  that is able to account both for the waving of the wavevector captured by Eq.~(\ref{eq:Qmodel}) as well as the magnetic phase diagram observed experimentally.

%magnetoelectric effect
In conclusion, we studied the magnetically ordered phases of a $4d$ cluster magnet GaMo$_4$S$_8$ by SANS, magnetization and magneto-current measurements. We found modulated magnetic states with sub-10\,nm periodicity that can be attributed to the stronger DMI due to the enhanced SOC of GaMo$_4$S$_8$, with respect to typical $3d$ transition metal based skyrmion hosts. The q-space distribution of the modulation vectors is markedly deformed, which is explained in terms of higher-order anisotropies becoming important in this $4d$ compound. In finite fields, a series of phase transitions is observed, which is assigned to the transformation of the cycloidal state to the SkL and a new single-q state. Moreover, these modulated spin textures are coupled to the ferroelectric polarization as evidenced by our magneto-current measurements. The exceptional stability of the modulated states against magnetic fields also indicates the importance of SOC in GaMo$_4$S$_8$. Our findings imply that a remarkable scaling down of the skyrmion size in bulk non-centrosymmetric materials can be achieved by exploring the plethora of $4d$ and $5d$ magnets.

%Notes: discussions should be extended by 1) connection between alpha and the micromagnetic parameters, 2) new skyrmion lattice state where the skyrmion tubes are modulated along the polar axis 

\section{Methods}
\textbf{Synthesis}
Single crystals were grown as described in Ref.~\onlinecite{Querre2014Electric}.

\textbf{Small-angle neutron scattering}
SANS experiments were performed at the Oak-Ridge National Laboratory High-Flux Isotope Reactor, using the General-Purpose Small-Angle Neutron Scattering Diffractometer \cite{Heller2018SANS,Wignall2012SANS}. A single crystal with a mass of $m=112$\,mg was mounted onto a rotatable sample stick with its $\left[1\bar{1}0\right]$ cubic direction parallel to the rotation axis. A neutron wavelength of $\lambda_n$=6\,$\angstrom$ with $\Delta\lambda_n/\lambda_n$=0.13 broadening was used with the detector set to a distance of 5\,m from the sample, employing a collimator of the same length.

\textbf{Magnetization and polarization measurements}
The magnetization measurements were performed using a Quantum Design MPMS. The longitudinal magnetic susceptibility calculated from static magnetization measurements as well as the magnetocurrent were measured at 10\,K while the field was rotated in fine steps within the (1$\bar 1$0) plane between successive field sweeps. The step size of the rotation was 1 and 2.5 degrees in case of the magnetocurrent and the magnetization measurements, respectively. Magnetocurrent (j=dP/dt) measurements were carried out using a Keysight Electrometer. The sample was contacted on parallel (111) faces, and correspondingly, the magnetic field induced changes in the electric polarization component parallel to the [111] direction was detected. The magnetoelectric susceptibility was calculated from the magnetocurrent by division through the constant magnetic field sweep rate of 1.2\,T/min. The field dependence of the magnetization was measured in decreasing temperature steps.

\section*{Data availability}
The data supporting the findings of this study are available from the corresponding author upon reasonable request.

\begin{acknowledgements}
We thank S.~Picozzi, and S.~Dong for useful discussions. This work was supported by the Deutsche Forschungsgemeinschaft (DFG) via the Transregional Research Collaboration TRR 80: From Electronic Correlations to Functionality (Augsburg-Munich-Stuttgart) and via the DFG Priority Program SPP2137, Skyrmionics, under Grant No. KE 2370/1-1. M.G. is supported by DFG through project A07 of SFB 1143 (project-ID 247310070), DFG projects No.~270344603 and 324327023. S.B. acknowledges support by National Research, Development and Innovation Office – NKFIH, FK 153003, Bolyai 00318/20/11 and by the BME-Nanonotechnology and Materials Science FIKP grant of EMMI (BME FIKP-NAT). This research was supported by the Ministry of Innovation and Technology and the National Research, Development and Innovation Office within the Quantum Information National Laboratory of Hungary. D.Sz. acknowledges the FWF Austrian Science Fund Grants No. I 2816-N27 and TAI 334-N. This research used resources at the High Flux Isotope Reactor, a Department of Energy (DOE) Office of Science User Facility operated by the Oak Ridge National Laboratory.

\end{acknowledgements}

\section*{Author contributions}
T.W., Y.T. and H.N. synthesized and characterized the crystal; Á.B., K.G., L.F.K. and L.B. measured magnetization; Á.B., D.Sz., and L.D.-S. performed the SANS experiments; K.G. performed angular dependent magnetization and polarization measurements; Á.B., I.K. and S.B. analyzed the SANS results; M.A. and M.G. developed the theory; Á.B., I.K., M.G. and S.B. wrote the manuscript with contributions from K.G.; I.K. and S.B. planned the project.

\section*{Competing interests}
The authors declare that there are no competing interests.

%\clearpage
%\input{GMS_SANS_prb_1.bbl}

%%%%%%%%%% Merge with supplemental materials %%%%%%%%%%
%\hline

\onecolumngrid
\hspace{8 cm}
\begin{center}
\textbf{\large Supplementary material: Squeezing N\'eel-type Magnetic Modulations by Enhanced Dzyaloshinskii-Moriya interaction of $4d$ Electrons}
\end{center}
\twocolumngrid
%%%%%%%%%% Merge with supplemental materials %%%%%%%%%%
%%%%%%%%%% Prefix a "S" to all equations, figures, tables and reset the counter %%%%%%%%%%
\setcounter{equation}{0}
\setcounter{figure}{0}
\setcounter{table}{0}
\setcounter{page}{1}
\makeatletter
\renewcommand{\theequation}{S\arabic{equation}}
\renewcommand{\thefigure}{S\arabic{figure}}
\renewcommand{\bibnumfmt}[1]{[S#1]}
\renewcommand{\citenumfont}[1]{S#1}
%%%%%%%%%% Prefix a "S" to all equations, figures, tables and reset the counter %%%%%%%%%%

\section{Differential susceptibility curves}
\begin{figure}[!ht]
	\centering
	\includegraphics[width=0.65\columnwidth]{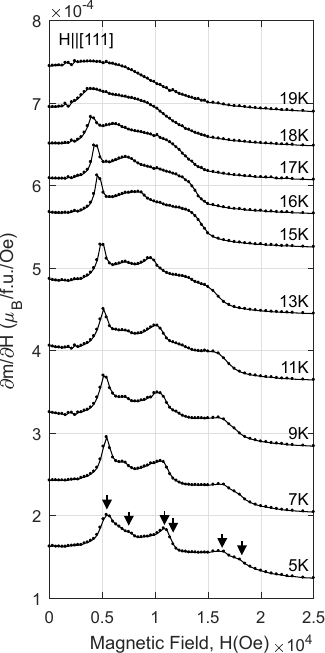}
	\caption{\textit{Differential susceptibility curves measured on 
			GaMo$_4$S$_8$ with the magnetic field applied along the [111] 
			direction. 
			The curves are shifted along the $y$ axis proportionally to the 
			temperature. The anomalies in the susceptibility curves, identified 
			as 
			phase transitions, are indicated by black arrows over the curve 
			measured at 
			$T=5$\,K. }}
	\label{fig:GMS_mH}
\end{figure}

\begin{figure}[!ht]
	\centering
	\includegraphics[width=\columnwidth]{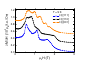}
	\caption{\textit{Differential susceptibility curves measured on 
			GaMo$_4$S$_8$ at 
			$T=5$\,K with the magnetic field applied along the main crystallographic axes. 
			The curves are shifted along the $y$ axis for better visibility.}}
	\label{fig:GMS_mH_angular}
\end{figure}

In Fig.~\ref{fig:GMS_mH} the differential susceptibility curves measured at various temperatures in the magnetic phases GaMo$_4$S$_8$, with the applied magnetic field parallel to the [111] crystallographic direction. The magnetic phase diagram in the main text displays the temperature-evolution of the anomalies in the susceptibility curves, identified as magnetic phase transitions. Differential susceptibility curves measured at $T=5$\,K along the main crystallographic axes are show in Fig.~\ref{fig:GMS_mH_angular}.  

\section{The effects of symmetrization and smoothing on the SANS image}

\begin{figure*}[!h]
	\centering
	\includegraphics[width=\textwidth]{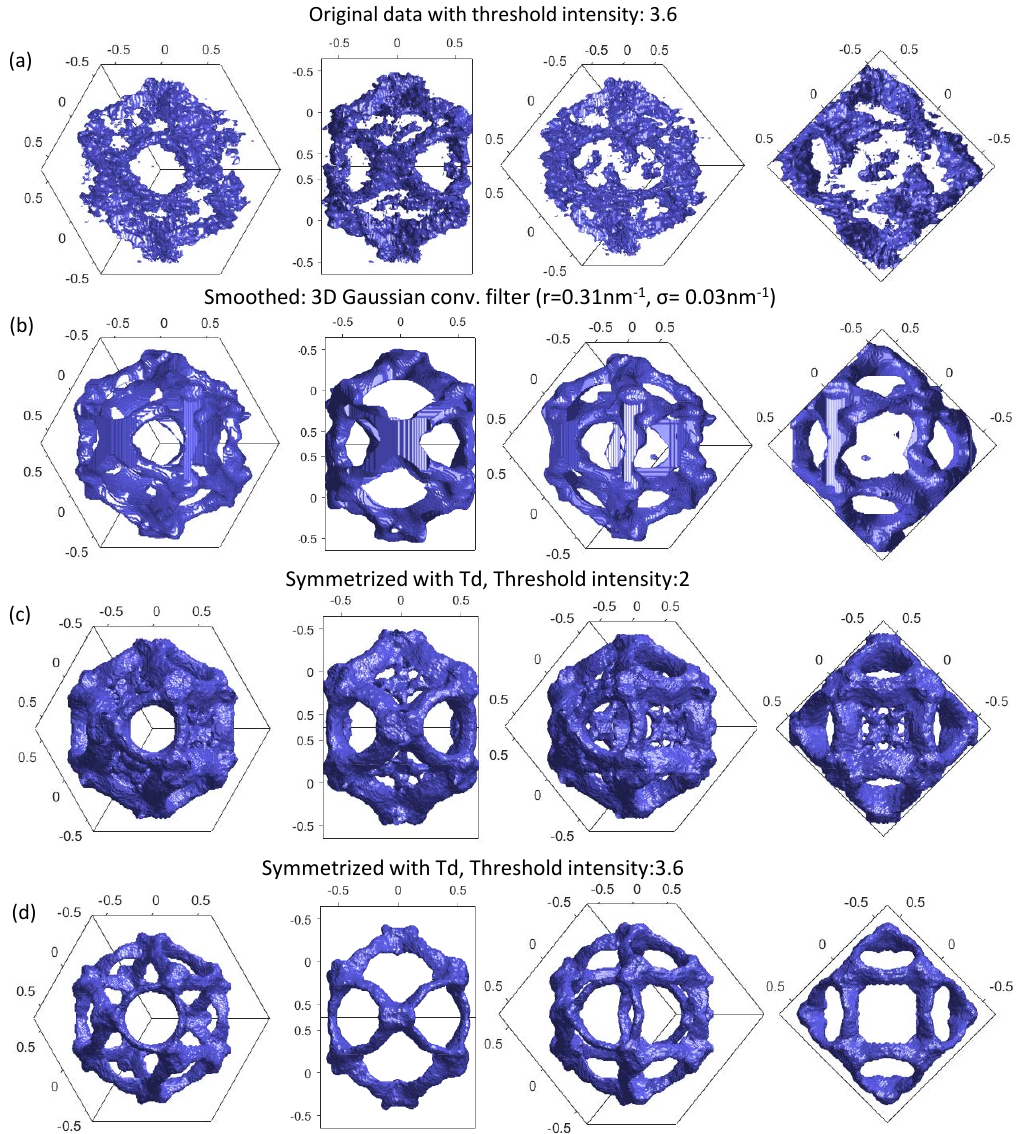}
	\caption{\textit{SANS reciprocal-space tomography images of the cycloidal 
	states of GaMo$_4$S$_8$ using different noise filtering methods. Panel (a) 
	shows the original image with a threshold intensity of 3.6 in arbitrary 
	units. In the second row (b), the same data is displayed using a strong 
	Gaussian convolutional filtering with a radius of $r=0.31$\,nm$^{-1}$ and a 
	FWHM of $\sigma=0.03$\,nm$^{-1}$. The effect of the symmetrization of the 
	reciprocal-space volume of the scattered intensity is demonstrated in rows 
	(c) and (d) with the visualization thresholds of 2 and 3.6 units, 
	respectively. The deflection of the rings from the 
	$\left\{111\right\}$-type planes is seen already on the raw data (a), 
	therefore it is not an effect of the symmetrization. The choice of the 
	intensity threshold affects the signal-to-noise ratio and the visualization 
	of the streaks within the $\left\{100\right\}$-type planes along the 
	$\left<100\right>$-type directions [c.f.~rows (c) and (d)].}}  
	\label{fig:GMS_RST_smoothing}
\end{figure*}

Figure \ref{fig:GMS_RST_smoothing} compares the three-dimensional structure of 
the magnetic $q$-vector distribution in GaMo$_4$S$_8$, reconstructed from the 
SANS wide-angle rotation experiments using different data manipulation methods. 
The sample was rotated around its $\left[1\bar{1}0\right]$ crystallographic 
axis (vertical axis in the figures) in 1$^{\circ}$ steps, with an acquisition 
time of 120\,s for each rotation angle. Prior to the reconstruction of the 3d 
image, the consecutive 2d scattering images were averaged over a 10$^{\circ}$ 
moving window to enhance statistics. The top row displays the three-dimensional 
$q$-distribution reconstructed from the scattering data after the angular 
averaging. The scattering intensities normalized to 1e9 monitoring counts are 
visualized above a certain threshold intensity value. The poor signal-to-noise 
ratio was improved by two distinct approaches.

Panel (b) shows the same reciprocal-space images after the pixel-wise 
application of a 3d Gaussian convolution filter, using a half-width of 
0.03\,nm$^{-1}$ with a radius of 31\,nm$^{-1}$. As a result, the four rings of 
scattering intensities are clearly visualized, further revealing their waving 
structure, deflecting from the $\left\{111\right\}$-type planes.

Another means to improve the signal-to-noise ratio of the scattering data is to 
exploit the symmetry properties of the crystal lattice. The statistics is 
enhanced through the averaging of the 3d scattering patterns connected by any 
of the symmetry operations of the crystal. In order to exclude any asymmetries 
introduced by the imbalance between the rhombohedral domain populations, we 
consider all the 24 symmetry operations of the high-temperature $T_{d}$ point 
group. The (c) and (d) rows in Fig.~\ref{fig:GMS_RST_smoothing} present the 
symmetrized images.

Whereas the image quality is greatly improved, the symmetrization apparently 
does not introduce artificial features to the images. A lower threshold 
intensity used for in row (c) results in broader rings and also visualizes the 
streaks observed in the $\left\{100\right\}$-type planes along the 
$\left<100\right>$-type directions. Such modulations may be introduced due to 
the boundary effects of the structural domain walls lying in the 
$\left\{100\right\}$-type planes on the magnetic patterns and is not discussed 
in the present study. Therefore, in the main text, the symmetrized 
visualization with a threshold intensity of 3.6\,a.u. Fig. 
\ref{fig:GMS_RST_smoothing} (d) is employed.  

\section{Angular dependence of the intensity}

\begin{figure*}[!h]
	\centering
	\includegraphics[width=\textwidth]{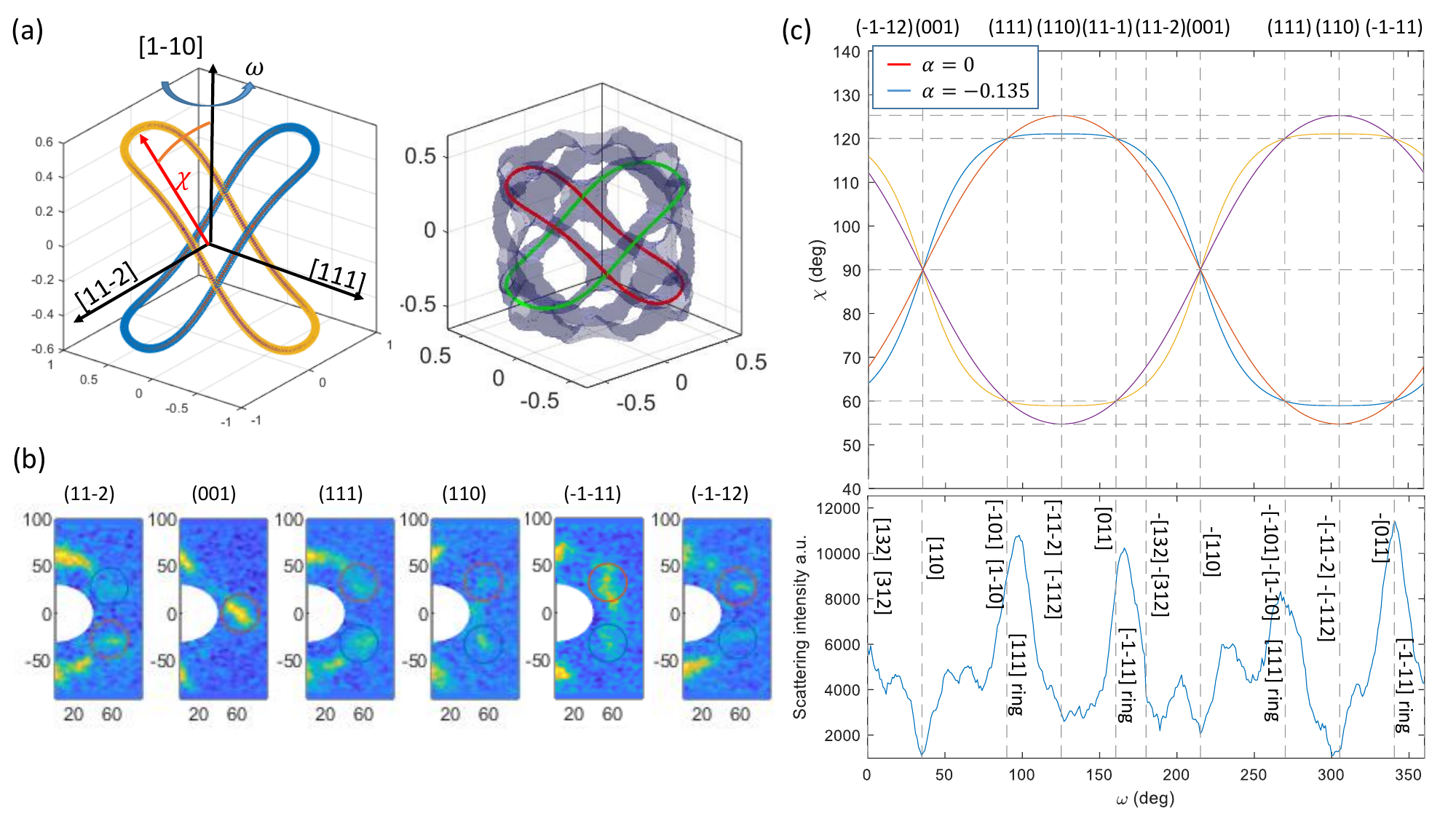}
	\caption{\textit{Angular dependence of the scattered intensity. The experimental frame of reference is shown in panel (a). Upon the rotation of the sample around its $\left[1\bar{1}0\right]$ axis by angle $\omega$ the intensity is integrated in regions parametrized by the angle $\chi$, where the rings are expected to intersect the image plane. The yellow and the blue rings corresponds to domains with rhombohedral axis parallel to $\left[\bar{1}1\bar{1}\right]$ and $\left[1\bar{1}\bar{1}\right]$, respectively. Detector images are shown for high symmetry directions in panel (b). The $\chi(\omega)$ calculated from the theory and the angular dependence of the intensity are shown in pnale (c).}}  
	\label{fig:GMS_Intensity}
\end{figure*}

The distributions of the scattered intensity over the rings are analysed as shown in Fig.~\ref{fig:GMS_Intensity}. Upon the rotation of the sample around its $\left[1\bar{1}0\right]$ axis by angle $\omega$ the intensity is integrated in regions parametrized by the angle $\chi$, where the rings are expected to intersect the image plane. The intensity is measured for rings in domains with rhombohedral axis parallel to $\left[\bar{1}1\bar{1}\right]$ and $\left[1\bar{1}\bar{1}\right]$, which directions are not perpendicular to the rotation axis, thus, non-trivial variation might be expected. Typical detector images for high symmetry directions are presented in Fig.~\ref{fig:GMS_Intensity} (b), which shows the areas where the intensity is summed. The $\chi(\omega)$ position of the q-vectors are calculated from the theory as discussed in the main text and they are plotted for zero and finite cubic anisotropy, $\alpha$ in Fig.~\ref{fig:GMS_Intensity} (c). 

Clear peaks appear in the angular dependence of the scattered intensity shown in Fig.~\ref{fig:GMS_Intensity} (c), when the scattering plane is perpendicular to the $\left[111\right]$ or the $\left[\bar{1}\bar{1}1\right]$ directions. The enhanced intensity at these positions is caused by the ring structures lying in the vicinity of the plane perpendicular to the respective rhombohedral axes. Beside this, no systematic variation of the intensity is observed for the symmetry equivalent $\left<1\bar{1}0\right>$ or $\left<11\bar{2}\right>$ directions.
 
\end{document}